\documentclass{article}
\usepackage{amssymb}

\usepackage{amsrefs}
\usepackage{amssymb}
\usepackage{amsmath}
\usepackage{graphicx}
\usepackage{color}
\usepackage{epsfig}
\usepackage[utf8]{inputenc}
\usepackage[T1]{fontenc}
\usepackage{epsfig}

\begin{document}

	\title{ Mechanics and Thermodynamics : \\ A link between the two theories}

	\author{ Henri Gouin \thanks{   
			E-mails:
			henri.gouin@univ-amu.fr; henri.gouin@ens-lyon.fr} }
	\date{\footnotesize Aix--Marseille University,
		IUSTI,  CNRS UMR 7343, Marseille, France.}
	\maketitle

	\begin{center}
	\date{\noindent\textit {Applications in Engineering Science 25 (2026) 100297}}
	\end{center}

	\maketitle
 
\begin{abstract}
	In this note, we analyze the relationships that should govern the use of thermodynamics in fluid mechanics in a way that we believe is understandable to mathematicians.
	We also aim to better define the reasons why mechanics and thermodynamics must be correctly linked by showing that the principle of virtual work expressed using a specific internal energy is perfectly suited to fluid mechanics problems, provided that a well-chosen  internal energy is proposed.
\end{abstract}
 
  \section{Introduction}
In the eyes of mathematicians, the first chapters of thermodynamics seem cluttered with an inextricable tangle of partial derivative calculations, placed like a thorny bush to block entry into this field. This detail, which physicists pursue like hunters drawn to their prey regardless of the thorns they encounter, stops them in  their tracks.\\ That is why, in the first part, we presented thermodynamic relationships using Poisson brackets, which are well suited to differential calculus.
In the second part, devoted to fluids, we reviewed the classical properties of internal energy, entropy, and temperature using our representation of Poisson brackets.
The concepts of heat quantity and calorimetric coefficients pose differentiation problems. We therefore revisited them in this context.
A third part introduces the concept of equilibrium based on the principle of virtual work applied to the specific internal energy of the fluid. It should be noted, however, that in the case of motion, fluid mechanics can sometimes appear to be independent of thermodynamics, as can be seen when vigorously stirring a spoon in a cup of coffee or propelling the cup to a high altitude without being able to heat it up. We also note that the principle of minimum   internal energy should be preferred to other principles such as that of minimum free energy, as the results obtained are generally not compatible.\\
Gibbs' thermodynamic surface allows us to study the properties of fluids geometrically and construct theorems that are comprehensible to mathematicians. We  study the stability of fluid equilibria in this geometry and justify the need for modified capillarity model reconsidered within the framework of the virtual work principle.   The specific internal energy used  is, for example, the origin of dispersive fluids.
Finally, we discuss the differences between  fluids and solids that must be taken into account in the development of new models.
 \section 
{Preliminary remarks on differential calculus} Consider a number of scalar    functions $(x, y, z, ...)$ at least twice differentiable  of a point $P$ on a two-dimensional manifold. Scalars $u$ and $v$ being two parameters, fixing the position of $P$, these different variables are functions of these two parameters. We  have taken three of these scalar functions
\begin{equation*}
\left\{
\begin{array}{l}
	\displaystyle 
dx  =  x^\prime_u\, du+ x^\prime_v\, dv \\  
	\displaystyle
dy  =  y^\prime_u\, du+ y^\prime_v\, dv  \\
\displaystyle
dz  =  z^\prime_u\, du+ z^\prime_v\, dv  
\end{array}\right. 
\end{equation*} 
By eliminating $du$ and $dv$ between these three equations, we obtain whatever values are given to $du$ and $dv$ (we will say more briefly whatever $d$) and whatever variables $x, y, z$ are chosen  
\begin{equation}
	[x, y]\, dz+ [y, z]\, dx+ [z, x]\, dy=0,\label{key1}
\end{equation}
where we have set 
\begin{equation}
	[x, y]=x^\prime_u\, y^\prime_v-x^\prime_v\,y^\prime_u\label{key2}
\end{equation}
Term $[x, y]$ is the \textit{Poisson bracket} of $x$ and $y$ with respect to the variables $u$ and $v$  \cite{Gouin_2020}.
Equation \eqref{key1} shows that if $[x, y]\neq 0$,
we can assign $x$ and $y$  the role of $u$ and $v$, i.e.  take them as parameters. A variable $z$  then has a differential
\begin{equation*}
	dz= a\, dx+ b\, dy\label{key3}
\end{equation*}
and noting that $[x,\, y]=-[y,\, x]$, we have  $\displaystyle a=\frac{[y,\, z]}{[y,\, x]}$ and  $b=\displaystyle\frac{[x,\, z]}{[x,\, y]}$.\\
In thermodynamics, we write $\displaystyle a =\left(\frac{\partial  z}{\partial x}\right)_y$ and
$\displaystyle b =\left(\frac{\partial  z}{\partial y}\right)_x$,
This means that $z$ is assumed to be calculated as a function of $x$  and   $y$, $a$ is the partial derivative $\displaystyle\frac{\partial  z}{\partial x}$, i.e. calculated assuming $y$ is constant. This is essential because if we expressed $z$ as a function of $x$ and an other variable  than $y$,   the partial derivative of $z$ with respect to $x$ would be different from the previous one and the partial derivative  gives the false idea of a ratio between two quantities $\partial z$ and $\partial x$ that do not exist.\\
The purpose of this paragraph is to show how we can give a more convenient notation.
The notation $\displaystyle \left(\frac{\partial  z}{\partial x}\right)_y$ is good, having on the one hand a precise location and on the other hand blocking well as a whole by its index  the false ratio $\displaystyle\frac{\partial  z}{\partial x}$. However, it is not very convenient in the sense that it does not lend itself to calculation. The best definition that can be given in another sense results from the equality \eqref{key2},
\begin{equation*}
\left(\frac{\partial  z}{\partial x}\right)_y =dz,\quad {\rm if}\quad dx=1\quad {\rm and}\quad dy=0.
\end{equation*}
This definition gives the practical calculation profile : $x, y, z$ being three functions of two arbitrary parameters to calculate $\displaystyle \left(\frac{\partial  z}{\partial x}\right)_y$, we must calculate $dx, dy$ and $dz$, then calculate $du$ and $dv$ using the two equations $dx=1$ and $dy=0$ and substitute these values into $dz$.
This is the calculation we performed in \eqref{key1}, which introduces Poisson's brackets  \eqref{key2}    and gives the result in the form of a true ratio :
\begin{equation}
	\left(\frac{\partial  z}{\partial x}\right)_y=\frac{[y,\, z]}{[y,\, x]}\label{key5}.
\end{equation}
The second member of this equality, combined with the definition of Poisson brackets, gives the value of $\displaystyle \left(\frac{\partial  z}{\partial x}\right)_y$ but does not yet constitute the notation that interests us. We specified above that a differential $d$ was defined when the values given to the differentials $du$ and $dv$ of the parameters were specified.
Now, for any differentiable scalar function, let $y$ correspond to a differential $d_y$, which will be obtained by giving $du$ and $dv$ the
values
\begin{equation}
	d_yu =-y^\prime_v \quad{\rm and}\quad     d_yv =+y^\prime_u.\label{key6}
	\end {equation}
	We then see that if $z$ is another function,
	$
	\displaystyle     d_yz = z^\prime_u \left(-y^\prime_v \right)+z^\prime_v\, y^\prime_u 
	$
	or precisely
	\begin{equation*}
		d_yz =\left[y,\,z\right].\label{key7}
	\end{equation*}
Poisson  bracket is thus simply a particular differential, and the partial derivative \eqref{key5} is then presented as the ratio  of two differentials,
\begin{equation*}
	\left(\frac{\partial  z}{\partial x}\right)_y=\frac{d_yz}{d_yx} \label{key8}
	\end {equation*}
We understand that the notation on the right, which is no more cumbersome than that on the left, is much preferable because it no longer forms a block and effectively corresponds to the ratio of two quantities $d_yz$ and $d_yx$ that exist independently of each other. Furthermore, it provides the fastest calculation method directly thanks to \eqref{key6}.
	Finally, the following properties allow the calculations to be performed without considering the chosen independent variables $u$ and $v$.
\subsection{Change of independent variables} We immediately verify that if, instead of the variables $u$ and $v$, we had taken two variables $u^\prime$ and $v^\prime$, all Poisson brackets would have been multiplied by the same quantity, and consequently the ratios would not have changed in value.  
If, in particular, we take two of the variables  $x$ and $y$, as independent variables, we will simply have $[x,\, y]=1$ or $d_xy=1$. Conversely, we can always set the Poisson bracket of two arbitrarily chosen variables equal to unity. 
\subsection{Poisson bracket relations}
We have the following formulas written either in differentials or Poisson brackets.
\begin{equation}
	\left\{
	\begin{array}{l}
	d_xx=0\quad\Longleftrightarrow\quad[x,\,x]=0\\  
		\displaystyle
		d_xy=-d_yx\quad\Longleftrightarrow
		\quad[x,\,y]=-[y,\,x] \\
		\label{Key26}
		\displaystyle
		d_xd_yz+d_yd_zx+d_zd_xz=0\quad\Longleftrightarrow
		\quad[x [y,\,z]] +[y [z,\,x]]+[z [x,\,y]]=0\\
			\displaystyle
		d_xd_yz =d_\alpha x\,\ {\rm with}\,\ \alpha = d_zy\quad\Longleftrightarrow
		\quad[[x,\, y],\,z]]  =-[z, [x,\,y]] \\
			\displaystyle
		d_xy\, d_t z+d_yz\, d_tx+d_zx\, d_t y=0 \,\ \Longleftrightarrow
		\;\ [x,  y] \, [t, z] +[y,  z] \, [t,  x]+ [z,  x] \, [t,  y] =0.
	\end{array}\right. 
\end{equation}
The third relation \eqref{Key26} is Jacobi's relation \cite{Gouin_2020}. It results from the relation $d[x,\,y]=[dx,\,y]+[x,\,dy]$.\\
Note   the following properties
\begin{equation*}
\frac{[z, x]}{[z, y]}\,\frac{[x, y]}{[x, z]}\,\frac{[y, z]}{[y, x]} =-1\qquad\Longleftrightarrow\qquad\	\left(\frac{\partial  x}{\partial y}\right)_z\left(\frac{\partial  y}{\partial z}\right)_x\left(\frac{\partial  z}{\partial x}\right)_y=-1.
\end{equation*}
Note also that if we consider two maps of a two-dimensional manifold, one obtained with the parameters $x$ and $y$ and the other with parameters  $z$ and $t$. A point $P$ is represented in these two maps by $P_1$ and $P_2$, respectively. If $d_xy =d_zt$,  the transformation $P_1\rightarrow P_2$ conserves closed surfaces. \\ Furthermore, if $a$ and $b$ are the partial derivatives of $z$ with respect to $x$ and $y$ ($dz= a\, dx+ b\, dy$), we have
\begin{equation}
	d_xa+d_yb =0\label{key9}
\end{equation}
\section{Application to thermodynamic calculations}
\subsection{Specific state, entropy, internal energy}
We  limit ourselves to the thermodynamics of a fluid whose state of a given mass can be defined by two quantities, one mechanical in nature, its specific volume $v$ (the inverse of which is the specific mass $\rho$), and the other thermal in nature, its specific entropy $\eta$ \cite{Rocard_1967}. We assume that each state corresponds to a specific internal energy $e$ and that knowledge of the characteristic function \cite{zemansky_1998}
\begin{equation}
	e=\varphi(v,\eta)\label{key10}
\end{equation}
is sufficient to determine, in all possible cases, the thermodynamic equilibrium of the fluid, but not its evolution. The various thermodynamic quantities that are introduced to study these equilibria, which are measured and which we will list,   can all be defined from $\varphi$ and its derivatives. These quantities are linked to each other by algebraic or differential relations that can be easily established from the previous remarks.
To a state of this fluid mass, we assign the geometric image of a point on a surface called the Gibbs thermodynamic surface, which is related to three coordinate axes where the values of $v, \eta$ and $e$ are plotted, its equation being \eqref{key10}. This surface has been constructed for a number of bodies. Knowledge of it summarises all the thermostatic properties of the body under study. Unfortunately, neither $\eta$ nor $e$ can be obtained experimentally, but only derived quantities.  
\subsection{Pressure, temperature}
The two quantities that are introduced first are {\it pressure and temperature}, defined by
\begin{equation*}
	de=-p\, dv+T\, d\eta,\quad{\rm or}\quad p=-\frac{d_\eta e}{d_\eta v},\,\ T =\frac{d_v e}{d_v \eta}, \label{key11}
\end{equation*}
where the quantities\,\ - $p$ and $T$ are  the partial derivatives of    $e$, we have, according to \eqref{key9}
\begin{equation}
	d_v p=d_\eta T\label{key12} 
	\end {equation}
	This formula, gives in new notation,  four   Maxwell's relations. They are obtained by dividing the two numbers in \eqref{key12} successively by the four Poisson brackets of the variables $v, p, \eta, T$ taken two by two; thus by dividing   by $d_v\eta$ we have
	\begin{equation*}
		\frac{d_v p}{d_v\eta}=\frac{d_\eta T}{d_v\eta}\quad {\rm or}\quad\left(\frac{\partial p}{\partial \eta}\right)_v =-\left(\frac{\partial T}{\partial v}\right)_\eta
	\end{equation*}
	Similarly, dividing by $d_vT, d_p\eta$ and $d_pT$, 
	\begin{equation*}
	\frac{d_v p}{d_vT}=\frac{d_T\eta}{d_T v},\quad 	\frac{d_p v}{d_p\eta}=\frac{d_\eta T}{d_\eta p},\quad 	\frac{d_p v}{d_p T}=-	\frac{d_T \eta}{d_T p},
	 \end{equation*}
	corresponding  in the classical notation to
 \begin{equation*}
 \left(\frac{\partial p}{\partial T}\right)_v =\left(\frac{\partial \eta}{\partial v}\right)_T,\quad\left(\frac{\partial v}{\partial \eta}\right)_p =\left(\frac{\partial T}{\partial p}\right)_\eta,\quad\left(\frac{\partial v}{\partial T}\right)_p =-\left(\frac{\partial \eta}{\partial p}\right)_T.
 \end{equation*}
\subsection{ Choice of variables}
It turns out that {\it we can measure} temperature and pressure, as well as specific volume, but we cannot measure entropy, at least not directly, which means that we do not like to use it as a variable.\\
Whether for this reason or others, the fact remains that the various thermodynamic quantities introduced are expressed as functions of two of the four variables $v, \eta, p$ and $T$.  
If we take the variables $v, p$, the state of the gas, for example, will be represented by a point on the plane relative to the axes $Ov, Op$; its evolution, represented by a curve on this plane, is the {\it Clapeyron diagram}. If we take $\eta, T$, we obtain the entropy diagram. Equality \eqref{key9} shows that when the evolution is cyclic, the areas of these two diagrams are the same, these areas representing a work. We can then set
\begin{equation*}
	d_vp = d_\eta T =1\label{key13}
\end{equation*}
To perform the calculations needed to convert from one system of variables to another, we need to know all the Poisson brackets of any two of the four variables, i.e. six Poisson brackets that are not zero a priori.\\
Two Poisson brackets are already known, equal to 1 : $d_vp = d_\eta T =1$. There are four remaining, but relation  \eqref{Key26}$^5$ tells us that they are not independent, as it is written here, 
\begin{equation}
	d_v\eta\,\, d_pT - d_p \eta\,\, d_vT =1\label{key14}
\end{equation}
The thermodynamic significance of this Mayer's relation will become clear once we have seen which thermodynamic quantities correspond to these Poisson brackets. Note that we only have {\it three independent ones}.  
When we say that there are three independent ones, it means that when we know the {\it numerical values} taken by three of them for a state of the system, we can deduce the {\it numerical value} of the fourth. Obviously, each of these Poisson brackets is a function of the state, which can be deduced from the single function $\varphi$ giving the specific internal energy, and   these functions are linked   to each other by  other differential relations \cite{Aifantis_Serrin_1983}. 
\subsection{Thermodynamic potentials}
Recall that we have the three Legendre transformed thermodynamic potentials of the specific internal energy $e=\varphi(v,\,\eta)$
\begin{equation}
	\left\{
	\begin{array}{l}
		{\rm Specific\ internal\ energy},\quad  e=e(v,\eta),\quad de=-p\, dv+T\,d\eta\\  
		\displaystyle
		{\rm Specific\ enthalpy},\quad  \tilde h=e+p\,v=\tilde{h}(v,\eta),\quad d\tilde h=v\, dp+T\,d\eta\\
		\label{Key15}
		\displaystyle
	{\rm Specific\ free\ energy},\quad  a=e-T\eta=a(v,T),\quad da=-p\, dv-\eta\,dT\\
		\displaystyle
	{\rm Chemical\ potential},\quad  g=e-T\,\eta+p\,v=g(p,T),\quad dg=v\, dp-\eta\,dT.
	\end{array}\right. 
\end{equation}
It is equivalent to characterise the gas under study by the data  of any of these  potentials. Depending on the problem being studied, i.e. the variables considered to be of interest, it will be advantageous to use one potential or another. It is clear that a potential must be given as a function of a mechanical variable, $v$ or $p$, and a thermal variable, $T$ or $\eta$. The data for a function   cannot   alone characterise a fluid.
\subsection{Equation of state   }
  The values $p, v, T$ are the three quantities that can be measured directly, whereas $\eta$ cannot, nor can any of the four potentials we have just listed. This is unfortunate because the measurements would give us directly the functions that fully characterise the body under study.\\ The measurements, on the other hand, give us directly the equation of state, the relationship between $T, p$ and $v$, i.e. f(v, p, T)=0.\\ 
  We can see that it is not possible to deduce the previous potentials from this, except with a great deal of arbitrariness. To produce them, for example the internal energy, we would need to solve the partial differential equation, 
    \begin{equation*}
    	f(v, -\varphi^\prime_v, \varphi^\prime_\eta)=0,
    \end{equation*}
    which is obviously possible, but which does not give us   an unique solution. In other words, bodies with very different thermodynamic properties can have 
    the same equations of state. 
    \\
    We therefore understand that it is necessary to study other quantities that can be measured, that these quantities  will naturally be known as a function of two of the three variables $p, v$ or $T$,   and that a whole set of differential calculations will be necessary if we want to deduce  from their knowledge the internal energy and entropy whose existence we have posited.

    \subsection{Calorimetric coefficients}
    During a change of state, the fraction $T\, d\eta$ of the change in internal energy is the {\it amount of heat received by the fluid};  it is written as $dQ$ in almost all treatises, which is confusing since it is not  a differential : it is only a differential form. In order not to stray from this unfortunate notation, some authors write it differently, such as $\tilde dQ$. This is what we will do.\\ Depending on the variables used to describe the state, we are led to define the various calorimetric coefficients $c, C, \ell, h, \lambda, \mu$  ($C$ is the specific heat at constant pressure, $c$ is the specific heat at constant volume),
    \begin{equation}
    	T\,d\eta = \tilde dQ = c\, dT+\ell\, dv =C\, dT+ h\, dp =\lambda \,dv+\mu\, dp,\label{key16}
    \end{equation}
    or by the definitions with $d_vp = d_\eta T =1$,
    \begin{equation}
    	c=T\, \frac{d_v\eta}{d_vT},\,\ C =T\, \frac{d_p\eta}{d_pT},\,\  \ell=\frac{T}{d_vT},\,\	    h=\frac{T}{d_pT},\,\ \lambda=-T\, d_p\eta,\,\ \mu = T\,d_v\eta. \label{key17}
    \end{equation}
  These six calorimetric coefficients are therefore directly given as a function of the four Poisson brackets of the variables, which we know, are not independent but linked by \eqref{key14}. Only three of these calorimetric coefficients are therefore independent.
  For example, between the first four we have the relationship
    \begin{equation*}
    	{d_p\eta}\,{d_vT}-{d_v\eta}\,{d_pT}=-1\,\  \Longleftrightarrow\,\ \frac{d_p\eta}{d_pT}-\frac{d_v\eta}{d_vT}=-\frac{1}{{d_vT}\,{d_pT}}\,\ \Longleftrightarrow\,\ C-c=-\frac{\ell\, h}{T}  
    \end{equation*}
    or between the last four, the relationship
    \begin{equation*}
    	{d_v\eta}\,{d_pT}-{d_p\eta}\,{d_vT}=1\,\  \Longleftrightarrow\,\ \frac{\mu}{h}+\frac{\lambda}{\ell}=1.
    \end{equation*}
    We are spoilt for choice when it comes to expressing Poisson brackets   as a function of the six calorimetric coefficients.
    These coefficients are naturally known as a function  of the three variables $p, v$ or $T$. What can we learn from knowing them?\\
    Suppose we already know the equation of state. We can see that $\ell$ and $h$ can be deduced from it and do not teach us anything new  (for example, if $p\,v= R\,T$ is the equation of state, $\ell=p$ and $h=-v$).  
    On the other hand, knowledge of one of the other four will give us one of Poisson  brackets $d_v\eta$ or $d_p\eta$ and then both thanks to \eqref{key14}. To clarify, let us assume that we know the equation of state,  i.e. $T=f(p, v)$ and the specific heat $c= m(p, v)$.\\ 
   Formulas \eqref{key17} give us $d_v\eta=\displaystyle \frac{c}{T}\,d_vT$, or $\displaystyle d_v\eta= m(p, v)\,\frac{f^\prime_p}{f}$  so $d_v\eta=\phi(p,v) $ and \eqref{key14} gives us $d_p\eta =\psi(p,v)$;\,\,\ so\,\,\ $\eta$ will be known to within a constant and \eqref{Key15}$^4$ shows that the chemical potential will be known to  {\it within a linear function of} $T$.\\ 
   We have no hope of reducing this arbitrariness that comes from integration constants. It is the third law of thermodynamics, or Nerst's principle, that will allow us to fix these integration constants  by postulating the properties of the  absolute zero. Unfortunately, no body is gaseous at absolute zero, and for solids, there are many other mechanical considerations to take into account. \\
   More generally, any three of the coefficients are known as a function of the {\it same two variables}; we will still know   four Poisson brackets and the same result will be achieved.
    \subsection{Relations  between calorimetric coefficients}
   Equations \eqref{key16} show us that $\displaystyle\frac{c}{T}$ and $\displaystyle\frac{\ell}{T}$, which are the partial derivatives of $\eta$ considered as a function of $T$ and $v$, satisfy a differential relation. Therefore, according to \eqref{key9}, we have
   \begin{equation*}
   	\label{Key18} d_T\frac{c}{T}+d_v\frac{\ell}{T}=0,\quad d_T\frac{C}{T}+d_p\frac{h}{T}=0,\quad d_v\frac{\lambda}{T}+d_p\frac{\mu}{T}=0.
   	\end {equation*}
   	Since $\ell$ and $h$ do not  intervene only $p, v, T$ (and can therefore be deduced from the equation of state), we see that $c$ and $C$  cannot be arbitrary functions if the equation of state is given. The first of these equations gives replacing $\ell$ with its value
    \begin{equation*}
    	\label{key19} d_T\frac{c}{T}+d_v\left(\frac{1}{d_vT}\right)=0\quad\Longleftrightarrow\quad d_Tc=-T\,d_v\left(\frac{1}{d_vT}\right), 
    \end{equation*}
    or  classically, $\displaystyle \left(\frac{\partial c} {\partial v}\right) _T=T\left(\frac{\partial^2p}{\partial T^2}\right)_v$.\\
   Thus, if the equation of state is known, for example, in the form $p=f(v,T)$ and if we are looking for $c$ as a function of $v$ and $T$, or $c=n(v,T)$, we will have
    \begin{equation*}
    	T f^{\prime\prime}_{v^2}-	n^\prime_v=0.
    \end{equation*}
    We know $c$; it is to an additive function   of $T$. This means that we only need to ask the experiments to tell us the variation of $c$ when $T$ varies alone, for an arbitrarily fixed value of pressure $p$.
    \\
    With the second equation \eqref{key17}, we similarly find
    \begin{equation}
    	\label{key20}  d_TC=T\,d_p\left(\frac{1}{d_pT}\right),\,\ {\rm or\  classically,}\,\  \displaystyle \left(\frac{\partial C} {\partial p}\right) _T=-T\left(\frac{\partial^2v}{\partial T^2}\right)_p. 
    \end{equation}
    However, we cannot derive anything further from the third equation \eqref{key17}, since $\lambda$ and $\mu$ involve $\eta$.
    \section{Application of thermodynamics to the study of fluid equilibrium}
   Thermodynamics alone does not provide a framework for studying a fluid in motion. Let us content ourselves with studying a fluid in equilibrium.\\
   The framework provided by thermodynamics is very simple. It is based on two concepts, those of position and energy, then on the assertion that energy is a function of position and on the principle that the equilibrium position is the one that minimises energy.\\
   To implement this model, we are invited to give an adequate representation of the position of the system, then to find which function of position is energy. This function will be characteristic of the system. The principle should then provide us with the equations of equilibrium.   \\
    For a fluid  considered as a thermo-mechanical system, the position has a mechanical aspect that we will specify. It has been given a thermal aspect;  this is what the second law of thermodynamics states.
    Thermodynamics treatises focus mainly on familiarising readers with the concept of entropy. With regard to the mechanical position of a fluid, the developments found in these treatises are insufficient for our study, as this position is always defined by a single number, the density $\rho$, which is assumed to be uniform (the same at every point in the fluid).\\
    In continuum mechanics, the position is represented mathematically by a diffeomorphism  $\mathcal F$, applying a reference space to the space occupied by the fluid at the instant  under consideration. This is what we  do,  not without noting that although this representation, already infinitely richer than the previous one, seems adequate for solids, it is not adequate for fluids because it excludes the well-known phenomena of mixing and diffusion. \\
    We know how to deduce from the knowledge of  $\mathcal F$ that of  deformation $\boldsymbol D$ and in particular of the specific volume $v$ at each point of the fluid. The total volume  occupied by the fluid in the domain ${\mathcal D}$ is\,\,\ $\displaystyle\mathcal U=\int_{\mathcal D} v\,dm$, where $dm$ is the mass element and the integral is a Stieltjes integral. 
    We will say that $\mathcal U$ is an additive quantity (this would not be true if there were diffusion or mixing).\\
    We assume --  {\it as a matter of principle} -- that energy and entropy are also additive variables, i.e. at every point in the fluid, there is an {\it  internal specific energy} $e$ and a {\it specific entropy}  $\eta$. We then have
   \begin{equation*}
   	\begin{array}{l}
   		{\rm Total\ volume}\quad \displaystyle\mathcal U=\int_{\mathcal D} v\,dm,\\
   		 {\rm Total\ entropy}\quad \displaystyle\mathcal S=\int_{\mathcal D} \eta\,dm,\\
   		 {\rm Total\ energy}\quad \displaystyle\mathcal E=\int_{\mathcal D} e\,dm.\\
   	\end{array}
   \end{equation*}
    The development of thermodynamic treatises invites us to consider that at each point $M_0$ of reference space, the internal specific energy is given by a characteristic function (i.e., characterising the fluid), 
    \begin{equation*}
    	e=\varphi(v, \eta, M_0)
    \end{equation*}
    We only consider the case where the fluid is heterogeneous $e=\varphi(v, \eta)$, a restriction that is not essential. All the thermodynamic quantities defined and studied above can be deduced from this function. Knowledge of this function allows us to write the equilibrium of the fluid using as a general principle, {\it the principle of virtual work}, which gives as a special case, for an isolated fluid, {\it the principle of energy extremum} \cite{Serrin_1959,Casal_1966,Marsden_Hugues_1994,Gouin_2007,Germain_2020}. 
   \subsection{Case of isolated fluid}
 That is, enclosed  in a fixed {\it adiabatic} chamber that does not allow any fluid volume or entropy to pass through. We therefore have for $\mathcal U $ and $\mathcal S$ given at
   to search among the fields $v$ and $\eta$ for the one that makes $\mathcal E $  extremum. If this field exists, a variation of this field $\delta v$, $\delta\eta$ such that $\delta\mathcal U =0$ and $\delta\mathcal S =0$ must give  $\delta\mathcal E =0 $.\\
   To express this condition of related extrema, we introduce two constant Lagrange multipliers  $-p_0$ and $T_0$ and write that
    \begin{equation*}
    	\delta (\mathcal E + p_0\, \mathcal U- T_0 \mathcal S ) =0\quad \Longleftrightarrow\quad\delta\int_{\mathcal D} (e+p_0\, v-T_0\, \eta)\, dm =0.
    \end{equation*} 
    We obtain :
    \begin{equation}
    	\varphi^\prime_v (v,\eta)=-p_0\,\,\ {\rm and}\,\,\   \varphi^\prime_{\eta} (v,\eta)=T_0, \label{keyA}
    \end{equation} 
  or $ p=p_0$  and  $T=T_0$.  The pressure and temperature are therefore uniform. We can always find an equilibrium for which $v$ and $\eta$ are uniform.\\ Then $\displaystyle v=\frac{\mathcal U}{\mathcal M}$, $\displaystyle \eta=\frac{\mathcal S}{\mathcal M}$ and $p_0$, $T_0$ can be deduced from \eqref{keyA} with $\displaystyle\mathcal M =\int_{\mathcal D}dm$.\\
    We will study the stability of this equilibrium production further on and see what other solutions there may be.
    
    \subsection{Case of fluid enclosed in a container where pressure and temperature are imposed}
    \subsubsection{Calculation}
    
    Let $T_0$ be the temperature of the walls. The container has a movable piston supporting a given external pressure $p_0$. Neither ${\mathcal U}$ nor ${\mathcal S}$ are imposed.\\
    If\,\ ${\mathcal U}$ increases by $\delta{\mathcal U}$, the exterior provides work $-p_0\,\delta{\mathcal U}$. If the entropy increases by $\delta{\mathcal S}$,  the exterior provides amount of heat $T_0\, \delta{\mathcal S}$ (of the physical dimension of a work).\\
    A variation of $\delta v$ and $\delta\eta$     increases the internal energy by $\delta{\mathcal E}$. The sum of virtual work is zero if this variation occurs from an equilibrium position. We write, as in the previous case,
    \begin{equation*}
    	\delta \mathcal E + p_0\,\delta \mathcal U- T_0 \,\delta\mathcal S   =0.
    \end{equation*}
    and we obtain the results given in \eqref{key20}. 
    The difference with the previous case is that here $p_0$ and $T_0$  are given but we arrive at the same conclusion.

    \noindent It can be seen that the use of the virtual work principle makes lengthy considerations of the various thermodynamic potentials unnecessary, since this principle introduces them automatically (in this subsection, we have therefore written that the chemical potential is extremum). Other cases are possible and would be treated in the same way with Lagrange multipliers, which in the case of a fluid can be interpreted as pressure or temperature. 
    \subsubsection{Internal energy and free energy}
\textit{It should be noted that some authors mistakenly consider the principle of minimum free energy to be equivalent to the principle of minimum internal energy. The proof is as follows}  :\\
    	Among the admissible positions, consider those at equilibrium with temperature $T$. The definition of free energy $a(v,T)$ shows that the equilibrium position is the one among the admissible positions that minimises
    	$$
    	\mathcal A=\int_{\mathcal D} a\,dm.
    	$$
    	We can therefore disregard the condition $\mathcal S$ and show that among the positions that give the same temperature as temperature at equilibrium, with only $\mathcal U$ being imposed, the equilibrium position minimises the free energy. 
    	We have dispensed with the knowledge of the total entropy, which is a number, but we have given ourselves a temperature {\it at every point}.\\
    	We then have the   principle: {\it For a given total fluid volume and temperature distribution, the free energy is minimum at equilibrium.}\\
    	(It is not clear how to impose the temperature, even if it is constant, but that is not the issue). Let us simply show that this principle does not give the same result as the   principle of internal energy extremum.\\
    	
    	{\it For free energy}: given $\mathcal U$ and $T$, 
    	\begin{equation*}
    		\mathcal A =\int_{\mathcal D} a(v, T)\,dm\,\  {\rm minimum}\quad \Longrightarrow\quad \delta(\mathcal A + p_{01}\,\mathcal U)=0 \quad {\rm where}\quad p_{01}={\rm Cst}. 
    	\end{equation*}  
    	By varying $\delta v$, we immediately obtain $p=p_{01}$.\\
    	
    	{\it For internal energy}: given $\mathcal U$ and $T$, 
    	\begin{equation*}
    		\mathcal E =\int_{\mathcal D} e(v, s)\,dm\,\,\,\   {\rm minimum} \quad \Longrightarrow\quad   \delta(\mathcal E +   p_{02}\,\mathcal U)=0 \quad {\rm where}\quad p_{02}={\rm Cst}.
    	\end{equation*} 
    	By varying $\delta v$ and $\delta\eta$, we obtain 
    	\begin{equation}
    		\delta e+p_{02}\, \delta  v=0\quad  \Longrightarrow\quad -p\, \delta v+ T\,\delta\eta + p_{02}\,\delta v =0 .\label{key23} 
    		\end {equation}
    	But $e_\eta^\prime (v, \eta)=T$ and $T\,\delta \eta =c\,\delta T+\ell\, \delta v$. Since $\delta T =0$,   substituting this result in \eqref{key23}, we obtain  $p=p_{02}+\ell$.\\
    		
    		Consequently, the two principles  give  {\it the same result only if} $\ell$ {\it is constant, which is not always the case}.  As long as the principle of least energy has not been rejected, we can use it because it is simple and allows us to examine the results it provides in order to compare them with experiments.

    \subsection{Case of a heavy fluid}
    The simple examples above give constant pressure and temperature in the fluid at equilibrium. This is no longer the case if the fluid is heavy, or more generally, if the fluid has an external potential $\Omega$ due to its position relative to the outside, which we have learned to consider from the study of pure mechanics.\\ The principle of virtual work invites us to vary {\it the position} and not just the specific volume, as was sufficient in the previous cases.\\
   The calculations are more delicate, but  they then indicate that we will have the conditions of equilibrium (see for example \cite{Germaina_1973})
    \begin{equation*}
    	v\,  {\rm grad}\, p = {\rm grad}\, \Omega\quad {\rm with}\,\ p=p_0 \,\ {\rm on\ the\ wall,}\,\  {\rm and\ always}\,\ T=T_0.\end{equation*}
   The pressure can therefore vary from one point to another in a fluid at equilibrium. On the other hand, the temperature $T$ is uniform. Therefore, neither $v$ nor $\eta$ are uniform.\\
   This well-known result leads us to make the following observation. Authors generally consider $v$ and $\eta$ as position variables, and $p$ and $T$ as stress variables. \ We see that there is no parallelism between $v$ and $\eta$; the mechanical position is not $v$ but $p$, of which $v$ is a derived quantity $\displaystyle\left(v ={\rm det} \,\frac{\partial M}{\partial M_0}  \right)$, where $M$ and $M_0$ are respectively the position of a particle in physical space and in reference space. 
   On the contrary, there is no reference to a thermal position quantity from which $\eta$ would be merely derived. There can be no question of imagining an external energy of the thermal type, and
   it is easy to understand the failures of attempts at  unitary theories where $\eta$ is considered as a position variable. Finally, it should be noted that the definition of entropy in statistical mechanics does not, at first glance, allow for any of these comparisons.\\
     Let us also say, although this goes beyond the limited domain in which we have deliberately placed ourselves in order to try to see things clearly, that when we apply thermodynamics to a chemical reaction, the position of the system is given by entropy and certain quantities $x, y,...$, which determine the physical state of the system.
      The internal energy is then a function $\varphi(\eta, x, y, ...)$, and we can deduce, using considerations analogous to the  previous ones, that the quantities 
     \begin{equation*}
     	T=\varphi^\prime_{\eta},\,\ X =\varphi^\prime_{x},\,\ Y=\varphi^\prime_{y},\,\ {\rm etc...}
     \end{equation*}
     curiously called thermodynamic {\it potentials}  are constant  at equilibrium.
    \\
   However, we have just seen, in the only example (of pure mechanics) that has been perfectly known for several centuries,  the corresponding {\it thermodynamic  potential}\,\ is pressure   and such a {\it potential  is not constant}. We pose the question without answering it. Do mechanical phenomena play a special role in the principles of thermodynamics? If so, where is this distinction found in the principles? If not, we must admit that the concepts of entropy (thermal position) and the {\it position}  of a chemical reaction, etc., are not yet sufficiently defined.
    \section{Study of the thermodynamic surface}
\begin{figure}[h]
   	\begin{center}
   		\includegraphics[width=14cm]{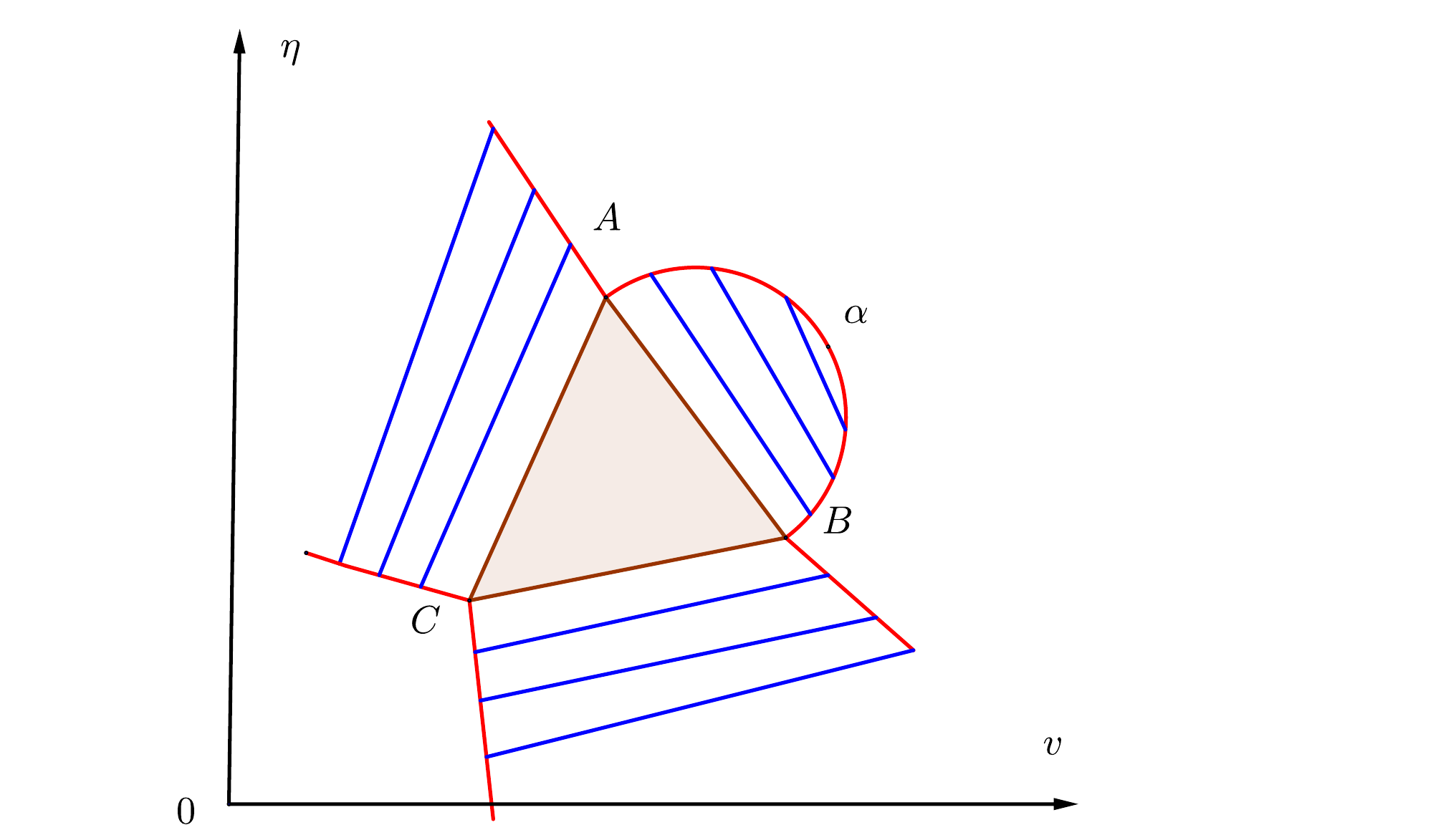}
   	\end{center}
   	\caption{{\protect\footnotesize {Surface $(\Im)$ projected onto the plane $Ov\eta$}}}
   	\label{1}
   \end{figure}

    Before studying the stability of an equilibrium position, let us make   geometric remarks about the thermodynamic surface $(\Im)$ of equation $e=\varphi(v,\eta)$. Since the function $\varphi$ is twice differentiable, the surface admits a tangent plane at every point whose normal has parameters $(p, -T, {\it 1})$. Since the parameters $p$ and $T$ are positive, the sections by planes $\eta=Cte$ are decreasing curves, while those by $v=Cte$ are increasing curves. Let $(\Im^\prime)$ be the convex hull of $(\Im)$ and $\mathcal O$ the open set of points of $(\Im)$ where the tangent plane has no other points in common with $(\Im)$ than the point of contact. The boundary of $\mathcal O$ is a curve $({\it\Gamma})$. The tangent plane to $(\Im)$ at a point $M_1$ of $({\it\Gamma})$ is then tangent at another point $M_2$ of $({\it\Gamma})$ (and generally only one other point). \\
   The segments $M_1M_2$ generate a developable surface $(\wp)$ which is a portion of the envelope of these {\it bitangent} planes  \cite{Kobayashi_Nomizu_1996}. Some of these planes may be tangent at a third point forming a triangle $ABC$ with the other two points.
   The developable surface $(\wp)$ is then formed of layers that connect to the triangle $ABC$ on one of its sides, their tangent planes being that of the triangle (see figure \ref{1}, projected onto the plane $Ov\eta$).
   The different sheets of $(\wp)$  end either at a triangle or at a critical point ($\alpha$ in the figure \ref{1})  where $M_1$ and $M_2$  coincide, or else they extend to infinity.\\
   The union of $\mathcal O$ and $(\wp)$ and the surface of the triangles constitutes the convex hull $(\Im^\prime)$ of $(\Im)$. The points of $(\Im^\prime)$ are either on $(\Im)$ or below $(\Im)$. The tangent plane  to $(\Im^\prime)$ varies continuously; it is always a tangent plane  to $(\Im)$  (we   avoid  the very exceptional cases where the  tangent plane to $(\Im)$ would be  at more than three points,  or even at an infinite number of points) \cite{Callen_1991}.
    \subsection{Position and equilibrium stability}
    Let us return to the first problem of isolated fluid equilibrium (where $\mathcal U$ and $\mathcal S$ are given). 
    We have seen that the fields $v$ and $\eta$  must be such that the pressure and  temperature are the same at every point  of the fluid (see \eqref{key20}).
    \\
    We  get the solution where $v$ and $\eta$ are uniform and equal to $\displaystyle   v_0=\frac{\mathcal U}{\mathcal M}$, $\displaystyle  \eta_0 =\frac{\mathcal S}{\mathcal M}$. 
    We say that the fluid is in equilibrium in a single phase, its pressure is $p_0=-\varphi^\prime_v (v_0,\eta_0)$ and its temperature is $T_0=\varphi^\prime_\eta (v_0,\eta_0)$.\\  Suppose that the point $M$ of $(\Im)$ corresponding to this equilibrium   is located in a non-convex part of $(\Im)$ and is therefore above $(\Im^\prime)$. The point $M^\prime$ of  $(\Im^\prime)$ with the same coordinates is located  either on a sheet of $(\wp)$ or in a triangle as $ABC$.
    
    $\bullet$ Let us consider the first case:
    $M^\prime$ is on the generatrix $M_1M_2$ whose coordinates are $(v_1,\eta_1)$ and $(v_2,\eta_2)$, and we can set
    \begin{equation*}
    	\left\{
    	\begin{array} {l}
    		\displaystyle
    		v_o= x\, v_1+ (1-x)\, v_2\\  
    		\displaystyle
    		\eta_o= x\, \eta_1+ (1-x)\, \eta_2
    	\end{array}\right.\quad {\rm with} \quad 0<x<1. 
    \end{equation*}
    At points $M_1$ and $M_2$   the tangent planes to $(\Im)$ are the same, and the states corresponding to fluid     have the same pressure and temperature,
    \begin{equation*}
    	\left\{
    	\begin{array}{l}
    		\displaystyle
    		p_1=-\varphi^\prime_v (v_1,\eta_1)=-\varphi^\prime_v (v_2,\eta_2)=p_2\\  \\
    		\displaystyle
    		T_1=\varphi^\prime_\eta (v_1,\eta_1)=\varphi^\prime_\eta (v_2,\eta_2)=T_2
    	\end{array}\right.\quad {\rm Generally}\,\ p_1\neq p_0\,\ {\rm and}\,\  T_1\neq T_0.
    \end{equation*}
    We can satisfy the equilibrium conditions by taking a part $x\mathcal{M}$ of the fluid  in state $(v_1, \eta_1)$  which  occupies a volume $\mathcal U_1$ and the remainder  part $(1-x)\mathcal{M}$ in state  $(v_2, \eta_2)$  which occupies a volume $\mathcal U_2$.\\
    We  have a {\it two-phase equilibrium}. \\
    The energy of such an equilibrium is $\mathcal E^\prime = x\,\mathcal{M}\, e_1+ (1-x)\,\mathcal{M}\, e_2$, or $E^\prime = \mathcal{M}\, e^\prime$, if $e^\prime$ corresponds to the ordinate of point $M^\prime$ of $(\Im^\prime)$.
    The two-phase equilibrium has an energy $\mathcal E^\prime$ lower than the energy $\mathcal E$ of the single-phase equilibrium. (A more complete but tedious discussion would show that no more stable equilibrium position can be found).
    If the point $M^\prime$ is located in a triangle, we  have a three-phase equilibrium. 
    
   $\bullet$ \textit{The second case}:   in the second problem, temperature and pressure are imposed, and the geometric problem is to establish the points of $(\Im)$ where the normal has a given direction. In this case, we can see in treatises of thermodynamics that there is {\it never a stable equilibrium under two phases}  (except in very exceptional cases).\\
   In classical treatises, we can read about all  considerations that can be made about multi-phase equilibrium. Our goal was  only to show that these questions are   consequences of the principle of virtual work.
   If phase 1 is denser than phase 2, we   call it liquid, and the other gas. The various positions of the curves $({\it\Gamma})$ are the curves of change of state. In fact, we can only speak of different types of states of matter as long as these curves separate the plane $(v, \eta)$ into disjoint domains, but, in fact, it is not the case for liquids and gases because these curves meet at a critical point.\\
   
   \noindent Note that the quantity $$L=T\,(\eta_2-\eta_1)$$   is the {\it heat of change of state} (heat of vaporisation,  fusion, etc.). On the corresponding layer of the developable surface, $p$ is a function of $T$. If we consider that $p$ and $T$ give the normal to the tangent plane that envelops $(\wp)$  and   of which $M_1M_2$ is the characteristic line,   we obtain  an original proof of Clapeyron's formula. \\
   \textit
   { The equation of the tangent plane is 
   	\begin{equation*}
   		(e-e_1)+p\,(v-v_1)-T\,(\eta-\eta_1)=0.
   	\end{equation*}
   Since $p=f(T)$, the tangent planes form a one-parameter family with characteristic tangent lines. The   developable surfaces are constructed by extending the tangent lines   of a curve  called edge of regression of the surface  \cite{Hilbert_2021}. The curve verifies 
   	\begin{equation}
   		\left\{
   		\begin{array}{l}
   			\displaystyle
   			(e-e_1)+p(T) \,(v-v_1(T))-T\,(\eta-\eta_1(T))=0\\ \label{Clapeyron}   
   			\displaystyle
   			-e_{1_T}^\prime-p \,v_{1_T}^\prime+T\,\eta_{1_T}^\prime+ \frac{dp}{dT}\,(v-v_1)-(\eta-\eta_1)=0 .
   		\end{array}\right. 
   	\end{equation}
   Relation $e_{1_T}^\prime=-p \,v_{1_T}^\prime+T\,\eta_{1_T}^\prime$  
   combined with second equation \eqref{Clapeyron} yields
   $
   \displaystyle T\frac{dp}{dT}\,(v-v_1)=T(\eta-\eta_1)
   $. Therefore,   we obtain Clapeyron's formula,}
   	\begin{equation*}
   		L= T (v_2-v_1)\frac{dp}{dT}.
   	\end{equation*}

    \noindent Other  change of phase theorems can be demonstrated. These results are the consequence of postulates relating to the existence of energy, entropy  and the principle of virtual work that governs equilibria.  They constitute a {\it synthesis}.
    \section
    {Introduction to Capillarity}
    
    Experiments show that equilibrium positions considered unstable can nevertheless be established in fluids
    (for example, in phenomena such as delayed boiling, supercooling, etc.).
    
    {\it The reason is simple} : these phenomena involve the concept of capillarity \cite{Rowlinson_Widom_2013}.\\
    In a two-phase equilibrium, the phases occupy volumes $\mathcal U_1$ and $\mathcal U_2$, separated by a surface $\it\Sigma$ (we will not consider more complex mixtures). 
    Experience shows the existence of a capillary energy proportional to the  surface of separation $\it\Sigma$ ($\mathcal E_c = \sigma\, \it\Sigma$ where $\sigma$ is the surface tension).
    This energy is added to the energy $\mathcal E^\prime$ to give a total energy  $\mathcal E^{\prime\prime} =\mathcal E^\prime+ \mathcal E_c$. If $\mathcal E^{\prime\prime}$ is greater than the equilibrium energy $\mathcal E$ of  a single phase   ($x$ must be close to 0 or 1),  the latter equilibrium will be stable (very weakly in fact), whereas we initially considered this equilibrium as unstable because $\mathcal E^\prime<\mathcal E$.\\
    Experiments show that equilibrium positions first    considered unstable can nevertheless be established in fluids.
    We will not examine these well-known results. We know that the choice of capillary energy is well verified by experiment. We even note that the total energy always remains, in accordance with the first principle, a function of the position of the fluid, since the surface is an element of this position.
    However, energy is then no longer an additive function relatively to mass, and there is no longer any {\it synthesis}.\\
    This {\it synthesis} may nevertheless remain, and the omission is not an error in the known principles, but stems from a flaw in reasoning, as the simple theory of   capillarity seems unrelated to the proposed model.

    The error was made when we posited that the specific energy $e$ defined in \eqref{key10} was a function of $v$  and $\eta$.
    In fact,  the treatises assume  that the fluid is uniform, its state  being given by the knowledge of the  density $v$ and entropy $\eta$; thus
    the energy  is necessarily of the form \eqref{key10}. It was too quickly assumed for a fluid mass whose state is given by two fields $v$ and $\eta$ that this was true for an infinitesimal element of the fluid.\\
    In general, at equilibrium, the internal energy $e$ is  not only  a function of the value of $v$ and $\eta$ but also of the values of their spatial derivatives of all orders at the point in question \cite{Vanderwaals_1895,Gouin_Sacco_2016,Dellisola_Gouin_Rotoli_1996}.\\
    Limiting ourselves to the first derivatives of these fields, i.e. $ {\rm grad} \,v$   and $ {\rm grad} \,\eta$, we must substitute a  richer  specific internal energy than \eqref{key10}  \cite{Casal_Gouin_1988b,Dunn_Rajagopal_1995}
    \begin{equation*}
    	e=\psi(v, \eta, {\rm grad} \,v,  {\rm grad} \,\eta).
    \end{equation*}
We may also assume that $e$ remains  invariant in an arbitrary rotation of the fluid and does not depend on its vectors except through their length  and angle.\\
It is clear that measurements made if $v$ and $\eta$ are uniform only allow the characteristic function  \eqref{key10} ;
\begin{equation*}
	\varphi(v, \eta)=\psi(v, \eta, \boldsymbol 0, \boldsymbol 0).
\end{equation*}
    It has been shown that term  $({\rm grad} \,v)^2$ in $e$, i.e.  
    $$e= \phi\left(v, \eta, ({\rm grad} \,v)^2\right)$$ 
    effectively account for the pressure differences prevailing on either side of the separation surface between the two fluids, and Laplace's theory can indeed be found  as  an {\it approximate theory} of a more refined model \cite{Garajeu_Gouin_Saccomandi_2013}. This model has the merit of being simple and  sufficient for common phenomena.  It  shows that capillary phenomena can be expressed in terms of internal volumetric energy and therefore in the statement of the principle of virtual work.\\ 
    Current considerations have now incorporated this type of internal energy model into the framework of dispersive fluids. It has been extended to dynamics by the fact that gradients can be expressed in space-time action and involve terms of the type $d\rho/dt$ by studying bubble liquids and other studies derived from them \cite{Gavrilyuk_Teshukov_2001, Ruggeri_Sugiyama_2021}.
    The consideration of terms in ${\rm grad} \,\eta$ has been little studied \cite{Casal_Gouin_1988b,Gouin_2017}. This case could arise, for example,  in interpreting temperature variations on either side of an interface or discontinuity surface, or in the study of  movements  with significant  temperature variations, such as in combustion phenomenon.
    Other cases could be considered for phenomena such as boiling, diffusion, mixing, turbulence  or cavitation \cite{Gavrilyuk_Liapidevskii_Chesnokov_2016,vergori_2008}, as these phenomena must be analysed within the framework of the principle of virtual action \cite{Gavrilyuk_Gouin_2020}.
   \\
   
  \emph{\textbf {Solids and fluids}}:
   The study may be extended to solids;   internal energy is then a function of the deformation   and the entropy fields.
   Instead of pressure, we have a stress tensor, and for capillarity, there are numerous coefficients to be exploited and measured \cite{rajagopal_1979}. \\  
   It  seems  that the theory is simpler for fluids than for solids.  The opposite is true, because in the case of fluids, we have completely ignored phenomena such as diffusion and mixing. The first property makes it appear simpler because the classical equations are indeed simpler than those of solids. However, if there is diffusion of matter or temperature, the position of a fluid is no longer, a priori, likely to be represented by a diffeomorphism between the reference space and the current space. It is not simply by phenomenologically adding a diffusion term that we obtain a coherent model.
    
   A drawing made by our ancestors on the walls of a cavern is still visible today. A glass of wine poured into water cannot be recovered a few seconds later.

	\bibliographystyle{amsrefs}
\bibliography{references}

\begin{bibdiv}
\begin{biblist}

\bib{Aifantis_Serrin_1983}{article}{
      author={Aifantis, E.C.},
      author={Serrin, J.B.},
       title={The mechanical theory of fluid interfaces and maxwell's rule},
        date={1983},
        ISSN={0021-9797},
     journal={Journal of Colloid and Interface Science},
      volume={96},
      number={2},
       pages={517\ndash 529},
  url={https://www.sciencedirect.com/science/article/pii/002197978390053X},
}

\bib{Callen_1991}{book}{
      author={Callen, H.B.},
       title={Thermodynamics and an introduction to thermostatistics},
   publisher={John Wiley},
        date={1991},
}

\bib{Casal_1966}{article}{
      author={Casal, P.},
       title={Principes variationnels en fluide compressible et en
  magn\'etodynamique des fluides},
        date={1966},
     journal={Journal de M\'ecanique},
      volume={5},
      number={2},
       pages={149\ndash 161},
}

\bib{Casal_Gouin_1988b}{article}{
      author={Casal, P.},
      author={Gouin, H.},
       title={Equations of motion of thermocapillary fluids},
        date={1988},
     journal={Comptes Rendus de l'Acad\'emie des Sciences de Paris, II},
      volume={306},
      number={2},
       pages={99\ndash 104},
}

\bib{Dellisola_Gouin_Rotoli_1996}{article}{
      author={Dell'Isola, F.},
      author={Gouin, H.},
      author={Rotoli, G.},
       title={Nucleation of spherical shell-like interfaces by second gradient
  theory: numerical simulations},
        date={1996},
     journal={European Journal of Mechanics - B/Fluids},
      volume={15},
      number={4},
       pages={545\ndash 568},
         url={https://hal.science/hal-00393974},
}

\bib{Dunn_Rajagopal_1995}{article}{
      author={Dunn, J.E.},
      author={Rajagopal, K.R.},
       title={Fluids of differential type: Critical review and thermodynamic
  analysis},
        date={1995},
        ISSN={0020-7225},
     journal={International Journal of Engineering Science},
      volume={33},
      number={5},
       pages={689\ndash 729},
  url={https://www.sciencedirect.com/science/article/pii/002072259400078X},
}

\bib{Gavrilyuk_Liapidevskii_Chesnokov_2016}{article}{
      author={Gavrilyuk, S.~L.},
      author={Liapidevskii, V.Yu.},
      author={Chesnokov, A.A.},
       title={Spilling breakers in shallow water: applications to favre waves
  and to the shoaling and breaking of solitary waves},
        date={2016},
     journal={Journal of Fluid Mechanics},
      volume={808},
       pages={441\ndash 468},
}

\bib{Gavrilyuk_Gouin_2020}{article}{
      author={Gavrilyuk, S.L.},
      author={Gouin, H.},
       title={Rankine–hugoniot conditions for fluids whose energy depends on
  space and time derivatives of density},
        date={2020},
        ISSN={0165-2125},
     journal={Wave Motion},
      volume={98},
       pages={102620},
  url={https://www.sciencedirect.com/science/article/pii/S0165212519303282},
}

\bib{Gavrilyuk_Teshukov_2001}{article}{
      author={Gavrilyuk, S.L.},
      author={Teshukov, V.M.},
       title={Generalized vorticity for bubbly liquid and dispersive shallow
  water equations},
        date={2001},
     journal={Continuum Mechanics and Thermodynamics},
      volume={13},
      number={6},
       pages={365\ndash 382},
}

\bib{Germaina_1973}{book}{
      author={Germain, P.},
       title={Cours de m\'ecanique des milieux continus},
   publisher={Masson},
        date={1973},
      number={vol.1},
        ISBN={9782225359378},
         url={https://books.google.fr/books?id=uQ4ZAQAAIAAJ},
}

\bib{Germain_2020}{article}{
      author={Germain, P.},
       title={The method of virtual power in the mechanics of continuous media,
  1 : Second-gradient theory},
        date={2020},
     journal={Mathematics and Mechanics of Complex Systems, translation from
  Germain, P. : La m\'ethode des puissances virtuelles en m\'ecanique des
  milieux continus, I. La th\'eorie du second gradient, Journal de M\'ecanique,
  {\bf 12} (1973), 235-274 and},
      volume={8},
      number={2},
       pages={153\ndash 190},
}

\bib{Gouin_2007}{book}{
      author={Gouin, H.},
       title={The d'alembert--lagrange principle for gradient theories and
  boundary conditions},
      series={Asymptotic Methods in Nonlinear Wave Phenomena, p.p. 79-95. Eds.
  T. Ruggeri and M. Sammartino and A.M. Greco},
   publisher={World Scientific},
        date={2007},
        ISBN={9789812707826},
         url={https://books.google.fr/books?id=y_poDQAAQBAJ},
}

\bib{Gouin_2020}{book}{
      author={Gouin, H.},
       title={Introduction to mathematical methods of analytical mechanics},
   publisher={Elsevier},
        date={2020},
        ISBN={9780128229866},
}

\bib{Gouin_Sacco_2016}{article}{
      author={Gouin, H.},
      author={Saccomandi, G.},
       title={Travelling waves of density for a fourth-gradient model of
  fluids},
        date={2016},
     journal={Continuum Mech. Thermodyn.},
      volume={28},
       pages={1511\ndash 1523},
}

\bib{Gouin_2017}{article}{
      author={Gouin, H.},
      author={Seppecher, P.},
       title={Temperature profile in a liquid-vapor interface near the critical
  point},
        date={2017},
     journal={Proceedings of the Royal Society of London. Series A,
  Mathematical and physical sciences},
      volume={473},
      number={20170229},
       pages={1\ndash 13},
         url={https://hal.science/hal-01492802},
}

\bib{Garajeu_Gouin_Saccomandi_2013}{article}{
      author={Gărăjeu, M.},
      author={Gouin, H.},
      author={Saccomandi, G.},
       title={Scaling navier-stokes equation in nanotubes},
        date={2013},
        ISSN={1089-7666},
     journal={Physics of Fluids},
      volume={25},
      number={8},
       pages={082003},
         url={http://dx.doi.org/10.1063/1.4818159},
}

\bib{Hilbert_2021}{book}{
      author={Hilbert, D.},
      author={Cohn-Vossen, S.},
       title={Geometry and the imagination},
   publisher={American Mathematical Soc.},
        date={2021},
      volume={87},
}

\bib{Kobayashi_Nomizu_1996}{book}{
      author={Kobayashi, S.},
      author={Nomizu, K.},
       title={Foundations of differential geometry, volume 1},
      series={John Wiley Publication in Applied Statistics},
   publisher={John Wiley},
        date={1996},
        ISBN={9780471157335},
         url={https://books.google.fr/books?id=BHp4wPTwaVwC},
}

\bib{Marsden_Hugues_1994}{book}{
      author={Marsden, J.E.},
      author={Hughes, T.J.R.},
       title={Mathematical foundations of elasticity},
   publisher={Courier Corporation},
        date={1994},
        ISBN={10: 0135610761},
}

\bib{rajagopal_1979}{article}{
      author={Rajagopal, K.~R.},
       title={A note on the drag for fluids of grade three},
        date={1979},
     journal={International Journal of Non-Linear Mechanics},
      volume={14},
      number={5-6},
       pages={361\ndash 364},
}

\bib{Rocard_1967}{book}{
      author={Rocard, Y.},
       title={Thermodynamique},
   publisher={Masson, Paris},
        date={1967},
}

\bib{Rowlinson_Widom_2013}{book}{
      author={Rowlinson, J.S.},
      author={Widom, B.},
       title={Molecular theory of capillarity},
      series={Dover Books on Chemistry},
   publisher={Dover},
        date={2013},
        ISBN={9780486317090},
         url={https://books.google.fr/books?id=eEDDAgAAQBAJ},
}

\bib{Ruggeri_Sugiyama_2021}{book}{
      author={Ruggeri, T.},
      author={Sugiyama, M.},
       title={Classical and relativistic rational extended thermodynamics of
  gases},
   publisher={Springer},
        date={2021},
      volume={197},
        ISBN={978-3-030-59143-4},
}

\bib{Serrin_1959}{book}{
      author={Serrin, J.},
       title={Mathematical principles of classical fluid mechanics},
   publisher={in Fluid dynamics 1, Ed. C. Truesdell, p.p. 125-263, Springer},
        date={1959},
}

\bib{Vanderwaals_1895}{article}{
      author={van~der Waals, J.D.},
       title={Archives n\'eerlandaises 28:121 (1895): The thermodynamic theory
  of capillarity under the hypothesis of a continuous variation of density},
        date={1979},
     journal={Journal of Statistical Physics, (Translation by J.S.
  Rowlinson),},
      volume={20},
      number={2},
       pages={197\ndash 200},
}

\bib{vergori_2008}{thesis}{
      author={Vergori, L.},
       title={Linear and nonlinear stability in non-standard theories of fluid
  dynamics},
        type={Ph.D. Thesis},
        date={2008, Universit{\`a} del Salento},
}

\bib{zemansky_1998}{misc}{
      author={Zemansky, M.~W.},
      author={Dittman, R.~H.},
      author={Scott, H.L.},
       title={Heat and thermodynamics},
   publisher={American Association of Physics Teachers},
        date={1998},
}

\end{biblist}
\end{bibdiv}
\end{document}